\documentstyle[aps,epsfig,prd,preprint]{revtex} 
\tightenlines
\def \beq{\begin{equation}}         \def \eeq{\end{equation}}
\def \beqa{\begin{eqnarray}}        \def \eeqa{\end{eqnarray}}
\def \bea{\begin{array}}	    \def \eea{\end{array}}
\def\plb#1#2#3{    {\it Phys. Lett. }{\bf B#1} (#2) #3}
\def\prd#1#2#3{    {\it Phys. Rev. }{\bf D#1} (#2) #3}

\def\prl#1#2#3{    {\it Phys. Rev. Lett. }{\bf #1} (#2) #3}

\def\zpc#1#2#3{    {\it Zeit. f{\"u}r Physik }{\bf C#1} (#2) #3}

\def\epj#1#2#3{{\it Eur. Phys. J.}{\bf C#1}(#2)#3}

\def\nomb{\nonumber}
\def \abs#1{\left| #1 \right|}
\def \lb{\left(}  \def \rb{\right)}
\def \fa{\sqrt{\frac{2}{3}}}    \def \fb{\sqrt{\frac{1}{3}}}
\def \fc{\sqrt{\frac{3}{2}}}    
\def \amp#1#2{#1_{#2}^{u,c} e^{i \delta_{#2}}}
\def \rk{i\frac{G_F}{\sqrt{2}}f_K F^{B\pi}_0(m_K^2)(m_B^2-m_\pi^2)}

\def \rkk{i\frac{G_F}{2}f_K F^{B\pi}_0(m_K^2)(m_B^2-m_\pi^2)}

\def \rpipi{i\frac{G_F}{2}f_\pi F^{BK}_0(m_\pi^2)(m_B^2-m_K^2)}
\begin{document}
\title{The interplay between Weak and Strong Phases and \\ Direct CP Violation 
           from the Charmless B-meson Decays }
\vskip -1cm
\author{{ Y.F. Zhou$^{a}$, Y.L. Wu$^{a,b}$, J.N. Ng$^{b,c}$, 
and C.Q. Geng$^{d}$
}\\
{\small $a$.\ Institute of Theoretical Physics, Chinese Academy of 
Science, Beijing 100080, China \\
$b$.\ National Center for Theoretical Sciences,
P.O. Box 2-131, Hsinchu, Taiwan 300\\
$c$.\ 
TRIUMF, 4004 Wesbrook Mall, Vancouver, BC, Canada V6T 2A3 \\
$d$.\ 
Department of Physics, National Tsing Hua University, 
Hsinchu, Taiwan 300\\
}}
%\date{}
\maketitle
\draft
\vskip -1cm
%\vskip -1cm
\begin{abstract}
\vskip -1cm

	 We  present a general analysis  on charmless 
B-meson decays $B\to \pi\pi$ and $\pi K$. It is noticed that the final 
state interactions and inelastic rescattering effects must be significant 
in order to understand the consistency of the current data. 
By using general isospin decompositions, the isospin amplitudes 
and the corresponding strong phases could be extracted from a global $\chi^2$
fit of the experimental
data. We emphasize that in general there are two, rather than one, relative 
strong phases in the decomposition. In the assumption of two equalstrong phases 
as considered in the literature, 
the current data , especially the ones concerning $B\to \pi^0 K^{0(\pm)}$decays,  will imply a 
large isospin amplitude $|a^c_{3/2}| > 40 $, which is larger by a factor of five 
than the one from the naive factorization estimation. When two different strong phases are considered, 
all the isospin amplitudes can become, within the $1\sigma$-level, comparable with the theoretical 
values. We also  show that the difference between the two strong phases cannot be too large 
and will be restricted by the most recent upper bound of $B\to \pi^0\pi^0$ decay. 
In any case, the strong phases are found to be large and 
the branching ratio of $B\to \pi^0\pi^0$ is likely to be enhanced by an order of magnitude 
in comparison with the one obtained from the naive factorization approach.   
Direct CP violations in all decay modes are also calculated and found to 
be close to the sensitivity of the present experiments.
\end{abstract}

%text
\newpage
 
\section{introduction}
%============>
Recently the  CLEO collaboration has reported measurements on  
the branching ratios of  rare hadronic B decays $B\to \pi\pi,\pi K$\cite{cleo,cleo-cp}. 
The data  have
attracted great interest from both theorists and experimentalists. 
The study of these channels will provide us
important insights on  understanding the effects of electroweak penguins 
(EWP) in B-system\cite{EWP},
and the final state interactions(FSI)\cite{hou,flei}, as well as extracting 
the weak CKM phase 
$\gamma=arg(V_{ud}V^{\ast}_{ub}/V_{cd}V^{\ast}_{cb})$\cite{gron98prl,nr,he,wz}. 
It may also open a window for probing new physics\cite{new-phy}.
%prob gamma, factorization, % cos gamma<0, large FSI phase, 

 From the current data, it is noticed that the branching ratio 
for $B\to \pi^+\pi^-$ is relatively small, $Br(B\to \pi^+\pi^-) 
\sim 4$ (in units of $10^{-6}$). The decays for $B\to \pi^+K^-,\
\pi^-\bar{K^0}$  have almost equal decay rates, 
i.e., $Br(B\to \pi^+K^-) \simeq Br(B\to \pi^-\bar{K^0}) \sim 17$. 
While an unexpectedly large branching
ratio for $B\to \pi^0\bar{K^0}$ decay was also observed, 
$Br(B\to \pi^0\bar{K^0})\sim 14$.
These measurements seem in conflict with the calculations 
based on the naive factorization hypotheses. 
For the first three decays, it was pointed out that the factorization 
approach may still be valid 
if one takes the weak phase $\gamma$ to be greater than $90^{\circ}$\cite{he}.
With such a large $\gamma$, i.e., $\cos\gamma < 0$, the interference between 
tree and penguin diagrams has opposite sign in $B\to \pi\pi$ and $B\to
\pi K$ decays. 
Thus the negative $\cos\gamma$ will suppress the decay rate for $B\to
\pi\pi$ and enhance that
for $B\to \pi^+ K^-$. As a consequence, the almost equal decay rates for
$B\to \pi^+K^-$ and $B\to \pi^-\bar{K^0}$ decay modes indicate the dominance of strong penguin.
However, the large rate for $B\to \pi^0 \bar{K^0}$
is not easily explained. Most recent analysis showed
that a large final state interaction (FSI) phase would be helpful to  enhance the branching ratio
for the $B\to \pi^0 \bar{K^0}$ decay\cite{hou}, but only considering the elastic-rescatterings remains 
insufficient to obtain the large central value of the data.

%model independent, isospin, su(3)
To understand the measured data, besides some model-dependent
calculations, a model-independent approach using a single relative strong phase 
has also been proposed for the study of $B\to \pi\pi, \pi K$
decays\cite{hou,pham,he}. Since the ordinary 
factorization approach suffers from  the uncertainties due to hadronic
matrix elements, such as 
the meson decay constants, B-meson form  factors and so-called effective
color number $N_c$. 
The model-independent analyses may be more useful as more data become available.
The approach based on the isospin $SU(2)$ and approximate flavor
$SU(3)$\cite{su3} symmetries 
of the strong interactions has been proposed to 
constrain \cite{fleischer,newbound} and extract the weak phase
$\gamma$\cite{nr}.  
It has been noticed that the ratios between the CP-averaged decay rates,  
such as $R = Br(B\to \pi^{\pm}K^0)/Br(B\to \pi^0K^{\pm})$ may provide us
important information 
on the weak phase $\gamma$. Most recently, it has been shown that the
weak phase $\gamma$ may 
be determined through three ratios among CP-averaged decay rates of $B\to
\pi^+\pi^-$, $\pi^+\pi^0$,
$ \pi^- K^+$ and $\pi^-K^0$ decays\cite{wz}. 
Where two solutions were obtained at the $1\sigma$ level, 
one with positive $\cos\delta$ and negative $\cos\gamma$, 
i.e., relative small strong phase $\delta$ 
and large weak phase $\gamma$, and another with negative $\cos\delta$ and
positive $\cos\gamma$.
The latter with positive $\cos\gamma$ seems to be favored by solutions
obtained from other constraints in 
the standard model but appears not to be as favorable as the one with
negative $\cos\gamma$ studies of all the 
existing charmless decays are taken in account.  However, 
there is still no complete  analysis in the
literature. 

% our work 
In this paper, we shall give a general analysis for all 7 decay
modes of $B\to \pi\pi$ and $\pi K$. For that purpose, we will start from a 
general model independent parameterization for all the 
decay amplitudes by considering both the isospin  
and simple diagrammatic decompositions.
We will show that there are in general 15 independent variables.
By assuming the SU(3) relations which appropriately account for 
SU(3) symmetry breaking effects, 
it allows us to reduce the 15 independent variables into 9. 
They consist of 6 isospin amplitudes 
and 2 relative strong phase as well as one weak phase $\gamma$. 
Note that once the relative strong phase is zero, $B\to \pi^-\bar{K}^0$ 
only receives contributions from penguin-type diagram and the amplitudes of 
the isospin $I=1/2$ and $I=3/2$ amplitudes from tree-type diagrams 
cancel each other, which provides an additional constraint\cite{he}. In fact, 
one of the isospin amplitudes becomes almost irrelevant due to the suppression factor of 
the CKM mixing element. With such a consideration, there are only 8 relevant unknown 
quantities with 5 isospin amplitudes and 3 phases. We show that the current 6 measured 
decay rates allow us to extract 6 unknown quantities as functions of two variable.  
The upper bound of the decay rate $B\to \pi^0 \pi^0$ also provides a bound for the 
difference between the two strong phases. Once taking the numerical value of the weak phase 
$\gamma$ to be the one obtained from other constraints in the standard model and 
fixing one of the strong phases, all the other parameters can be determined. With these 
determined parameters, we are then able to predict the branching ratio of the 
$B\to \pi^0\pi^0$ decay mode which is yet unmeasured due to the difficulty of its identification 
by the current detector. In addition we also present predictions for direct CP violations 
in all 7 decay channels of 
$B\to \pi\pi, \ \pi K$. In our numerical fitting, we have adopted the $\chi^2$ 
analysis for the CLEO data in order to have a systematic treatment on the experimental 
errors. 
  
	In general, according to the Watson theorem, there are two independent relative 
strong phases associating with the isospin amplitudes. They are often assumed 
to be equal in the literatures\cite{hou,he,pham}. 
In this work, we shall make a more general analysis with two relative strong phases.
 It is   shown that the equal phase assumption will result
in large enhancement of isospin amplitude  $a^c_{3/2}$ which will be 5 times larger
than  the one calculated from the factorization approach. The value of the strong phase is 
found to be $\delta \simeq \pm 95^{\circ}$. These large values may imply large
inelastic FSI or indicate the possible new physics effects.  However, if the two strong phases
are different, the value of $a^c_{3/2}$ can be lower and is comparable with the 
usual factorization calculations.

	 It is remarkable to observe that within $1\sigma$ all the 6 decay rates can be
consistently fitted for a large range of the weak phase $0^{\circ}<\gamma<
180^{\circ}$ for the above two cases.  It is also of interest to note that one of isospin amplitudes 
and the strong phases have a weak dependence on 
the weak phase $\gamma$. Three isospin amplitudes show a moderate dependence 
on the weak phase $\gamma$. Only one isospin amplitude is sensitive to  
$\gamma$.  In particular, the fitting values for the 4 usual isospin amplitudes  
considered in most of the literature could still be comparable with the ones obtained 
by using naive factorization  approach. The resulting large strong phases 
may be regarded as a strong indication of large FSI in $B\to \pi\pi,\pi K$ decays. 

%the additional new isospin 
%amplitude must  be larger by a factor of 7-9 than the one estimated from
%the naive factorization approach, 
%but its absolute value remains small and is only about $5\sim 10\%$ of
%the largest isospin amplitude when the phase $\gamma$ varies from 
%$0^{\circ}-180^{\circ}$. 
%It may not be  difficult to understand such a significant enhancement for
%the additional new isospin amplitude when  considering possible large effects 
%from the inelastic rescatterings such as:
%$B\to D+D \to \pi + \pi$ and $B\to D + D_s \to \pi + K$. 
%The resulting large strong phase $\delta 
%\simeq 95^{\circ}$ also indicates possible large inelastic rescattering effects. 

\section{The general framework}
 We begin with writing the decay amplitude of $B\to \pi\pi, \pi K$ in the following general form:
\beq
A^{\pi\pi(\pi K)}= \lambda^{d(s)}_{u} A^{\pi\pi(\pi K)}_u+\lambda^{d(s)}_c 
                  A^{\pi\pi(\pi K)}_c,
\eeq
where $\lambda^{d(s)}_{u}=V_{ub}V^*_{ud(s)}$ and 
$\lambda^{d(s)}_{c}=V_{cb}V^*_{cd(s)}$ are the products of CKM matrix elements.
The term proportional to $\lambda^{d(s)}_{t}=V_{tb}V^{\ast}_{td(s)}$ has been absorbed into 
the above two terms by using the unitarity relation, 
$V_{ub}V^*_{ud(s)}+V_{cb}V^*_{cd(s)}+V_{tb}V_{td(s)}=0$.

we also find it useful to adopt the isospin decomposition for the decay amplitudes 
\beqa
A^{u,c}_{\pi^-\pi^+}&=&\fa \amp{a}{0} +\fb \amp{a}{2} \label{su2-1}\\
A^{u,c}_{\pi^0\pi^0}&=&\fb \amp{a}{0} -\fa \amp{a}{2} \\
A^{u,c}_{\pi^-\pi^0}&=&-\fc \amp{a}{2}  \\
A^{u,c}_{\pi^+K^-}&=&\fa \amp{a}{1/2} +\fb \amp{a}{3/2} \\
A^{u,c}_{\pi^0\bar{K^0}}&=&\fb \amp{a}{1/2} -\fa \amp{a}{3/2} \\
A^{u,c}_{\pi^0K^-}&=&-\fc \amp{a}{3/2} - \frac{1}{\sqrt{2}}A^{u,c}_{\pi^-\bar{K^0}} 
\label{su2-6}
\eeqa
with 
\begin{equation}
A^{u,c}_{\pi^-\bar{K^0}}  = \fa b^{u,c}_{1/2}e^{\delta'_{1/2}} - \fb \amp{a}{3/2} 
\end{equation}
	Where $a_I^{u,c}$ and $b^{u,c}_I$  are the isospin amplitudes and $\delta_I$ 
and $\delta'_{1/2}$ are the strong 
phases due to final state interactions. In some literatures the strong phase of isospin amplitude
$b^{u,c}_{1/2}$ is assumed to be equal to the one of $a^{u,c}$ for simplicity\cite{hou,he,pham}. 
However, in the most general case, these strong phases are not necessarily the same, since
they arise from the effective Hamiltonian with different isospin.  
The subscript $I=0,\ 2,\ 1/2,\ 3/2$ denote the isospins of the amplitudes.
The advantage of the isospin decomposition allows one to use $SU(3)$ relations 
including leading order $SU(3)$ breaking effects. In other words, the isospin amplitudes are assumed to
satisfy the following relations:
\beqa
a^{u,c}_{0}&\simeq &(f_\pi/f_K) a^{u,c}_{1/2}, \ \ \ a^{u,c}_{2}\simeq (f_\pi/f_K) a^{u,c}_{3/2} \nomb\\
\delta_{0}&\simeq &\delta_{1/2},  \ \ \  \delta_{2} \simeq \delta_{3/2}
\eeqa
where $f_\pi$ and $f_K$ are the $\pi,K$ meson decay constants with
$f_\pi/f_K\simeq 0.8$. For convenience, we define two phase differences as follows:
\beqa
\delta  &=&\delta_{3/2}-\delta_{1/2} \nomb \\
\delta' &=&\delta'_{1/2}-\delta_{1/2}
\eeqa
 
Practically, the decay amplitudes are evaluated by calculating various Feynman diagrams.
 In order to see how those isospin amplitudes receive contributions from diagrams, 
we also present a simple diagrammatic decomposition. The diagrams can in general be classified 
into 6 types denoted by $T$(tree), $C$(color suppressed tree)
$P$(QCD penguin), $P_{EW}$(electroweak penguin) and $P^C_{EW}$( 
color suppressed electroweak penguin) \cite{su3new}:
\beqa
A_{\pi^-\pi^+} &=& T+P+{2\over 3}P^C_{EW} \\
A_{\pi^0\pi^0} &=& \frac{1}{\sqrt{2}}(-C+P-P_{EW}-{1\over 3}P^C_{EW})\\
A_{\pi^-\pi^0} &=& \frac{1}{\sqrt{2}}(-T-C-P_{EW}-P^C_{EW}) \\
A_{\pi^+K^-}   &=& T'+P'+{2\over 3}P'^C_{EW}\\
A_{\pi^0\bar{K^0}}   &=& \frac{1}{\sqrt{2}}(-C'+P'-P'_{EW}-\frac{1}{3}P'^C_{EW}) 
\eeqa
where the primed and unprimed quantities are the amplitudes in $B\to \pi K$
and $B\to \pi \pi$ decays. They  roughly differ by a factor $f_\pi/f_K\simeq 0.8$
when the $SU(3)$ flavor symmetry breaking effects are considered.  

Combining the two decompositions, it is straight forward to get the following relations:
\beqa
\amp{a}{1/2}&=&\frac{1}{\sqrt{6}}\lb 2T'-C'+3P'-P'_{EW}+P'^C_{EW}\rb^{u,c}\ , \\
\amp{a}{3/2}&=&\frac{1}{\sqrt{3}}\lb T'+C'+P'_{EW}+P'^C_{EW} \rb^{u,c} \label{a2c}\ .
\eeqa
        From the above equations, one may easily see the relative magnitudes among those 
$SU(2)$ invariant amplitudes. If the inelastic rescattering effects are small,
$T',C'$ will only contribute to the term proportional to $\lambda^s_u$. Therefore
one may expect that $T'(C')^u \gg T'(C')^c$. This will lead to 
$a^u_{1/2(3/2)}\gg a^c_{1/2(3/2)}$. Since $a^c_{1/2}$ receives contributions
from QCD penguins, while $a^c_{3/2}$ only gets contributions from EWPs, one may conclude that
$a^c_{1/2} \gg a^c_{3/2}$.

To obtain relations for the isospin amplitude $b^{u,c}_{1/2}$, one needs to be careful in adopting the 
diagrammatic decomposition implied by the naive factorization ansatz. This is because 
the resulting relative strong phase is zero in the factorization approach , i.e., 
$\delta =\delta_{3/2}- \delta_{1/2} =0$ and $\delta'=0$. As a consequence, 
\begin{eqnarray}  
A_{\pi^0K^-}   &=& \frac{1}{\sqrt{2}}(-T'-C'-P'-P'_{EW}-\frac{2}{3}P'^C_{EW})\\
A_{\pi^-\bar{K^0}}   &=& P'-{1\over 3}P'^C_{EW} \label{penguin}
\end{eqnarray}

The amplitudes with isospin $I=1/2$ and $I=3/2$ from tree-type graphs cancel each other
in Eq.(\ref{penguin}). 
Thus the total amplitude only receives contributions from penguin diagrams in this case, namely
\begin{equation}
A^{u,c}_{\pi^-\bar{K^0}}  = \fa b^{u,c}_{1/2} - \fb a^{u,c}_{3/2} 
= \lb P'-\frac{1}{3}P'^C_{EW} \rb^{u,c}
\label{b1/2}
\end{equation}

  If  assuming the $t$-quark dominance in the
penguin loops, one finds from  eq.(\ref{b1/2}) that : 
\beq
A^{u}_{\pi^-\bar{K^0}}  \simeq A^{c}_{\pi^-\bar{K^0}}  \quad or \quad  
b^u_{1/2}\simeq  b^c_{1/2} + \frac{1}{\sqrt{2}} a^{u}_{3/2} - \frac{1}{\sqrt{2}} a^{c}_{3/2} .
\label{approx}
\eeq
which may be assumed for simplicity to be approximately valid after considering final 
state interactions with nonzero strong phases. In the numerical calculations, 
we have check that the amplitude $b^u_{1/2}$ 
is less important due to the strong suppression of the CKM factor (for instance, 
even if taking $b^u_{1/2}\simeq  b^c_{1/2}$, the results remain almost unchanged).
   
%dicussion on delta_2 , a2c
  With the above analyses, let us provide an intuitive discussion on 
how to yield a large branching
ratio for $B\to \pi^0\bar{K^0}$ decay by
appropriately  choosing the isospin amplitudes. Note the fact that
$\lambda^s_u \ll \lambda^s_c$, 
one may roughly estimate  the ratio between  $Br(B\to \pi^0\bar{K^0})$ and
$Br(B\to \pi^+K^-)$ by 
neglecting the terms containing the CKM factor $\lambda^s_u$ 
\beqa
R=\frac{Br(B\to \pi^0\bar{K^0})}{Br(B\to \pi^+K^-)}&\simeq&
\abs{\frac{\fb a^c_{1/2}-\fa a^c_{3/2}e^{i\delta}}
         {\fa a^c_{1/2}+\fb a^c_{3/2}e^{i\delta}}}^2\,, 
\label{R}
\eeqa 
 with $\delta=\delta_{3/2}-\delta_{1/2}$. Once neglecting $a^c_{3/2}$, the
ratio may be  
 simply given by $R\simeq \frac{1}{2}$, which is much smaller than the
central value of the data
 $R = 0.84 $. It indicates that to enhance the decay rate of $B\to
\pi^0\bar{K^0}$, the isospin 
 amplitude $a^c_{3/2}$ should not be neglected. Its small value may
provide a sizable contribution 
 for a large value of $\delta>\pi/2$. This is because in this case there
exists a 
 constructive interference between $a^c_{1/2}$ and $a^c_{3/2}$ in $B\to
\pi^0\bar{K^0}$ 
 and a destructive interference in $B\to \pi^+K^-$. The situation is
quite similar to 
 the case for a large $\gamma>\pi/2$, which is considered to enhance $B\to
\pi K $ and 
 decrease $B\to \pi\pi$ decay rates. From eq.(\ref{R}), it is easily
seen that 
the value of $a^c_{3/2}$ satisfies
\beqa
 a^c_{3/2} \geq \frac{\sqrt{2R}-1}{\sqrt{2}+\sqrt{R}}a^c_{1/2} \simeq 0.12
\ a^c_{1/2}\,.
\eeqa  

  With the above considerations,  there are only 8 relevant quantities:
 $a^{u,c}_{1/2},\ a^{u,c}_{3/2},\ b_{1/2}^c,\ \delta,\ \delta',\ \gamma $,
 which should be constrained by six measured decay rates and one upper bound. 
When taking the weak phase $\gamma$ as a free parameter, the
rest 6 variables can be determined from 6 equations of eqs.(\ref{su2-1}-\ref{su2-6}). As
 the errors in the current data remain considerable large, one may not
take the central values 
 of the data to be too serious. Thus by only using the central values of
the data to determine the 
 6 variables may not be good enough. To take into account the experimental
errors in a systematic way,
 we shall adopt a global $\chi^2$( least square) analysis for the present
data. This treatment allows us
 to obtain not only the central value but also the errors for the fitting
amplitudes.   
 Our fitting will be carried out  by using the standard $\chi^2$ analysis
program package
MINUIT\cite{cernlib}.

% discussion of fitting
\section{Results and discussions}

         In order to compare with the values estimated from the factorization, it is
necessary to explicitly see how large of the isospin amplitudes $a^{u,c}_{1/2,3/2}$
and $b_{1/2}$ from the factorization calculations, we present the relevant formulae for
the $B\to \pi\pi,\pi K$ decay amplitudes with the assumption of factorization\cite{lu}:        
\beqa
 A_{\pi^+K^-}&=&\rk \lb \lambda^s_u
a_1-\lambda^s_t(a_4+a_{10}+(a_6+a_8)R_4)\rb \\
 A_{\pi^0\bar{K^0}}&=&-\rkk \lambda^s_t \lb
a_4+a_6R_5-\frac{1}{2}(a_{10}+a_{8}R_5)\rb\nomb\\
                  && -\rpipi \lb \lambda^s_u a_2-\lambda^s_t
\frac{3}{2}(a_9-a_7)\rb\\
A_{\pi^0K^-}&=&-\rkk \lb \lambda^s_u a_1-
              \lambda^s_t(a_4+a_6R_4+(a_{10}+a_{8}R_4)\rb\nomb\\
                  && -\rpipi \lb \lambda^s_u a_2-\lambda^s_t
\frac{3}{2}(a_9-a_7)\rb \\
 A_{\pi^-\bar{K^0}}^c &=&-\rk \lambda^s_t \lb
a_4+a_6R_5-\frac{1}{2}(a_{10}+a_8R_5)\rb 
\eeqa   
where  $f_{\pi,K}$ and $F^{B\pi,BK}$ are the decay constants and $B$-meson
form factors respectively, $R_4=2 m^2_K/(m_b-m_u)(m_u+m_s)$ and $R_5=2
m^2_K/(m_b-m_d)(m_d+m_s)$. In the flavor $SU(2)$ limit, one has 
$R_4\simeq R_5\simeq 2 m^2_K/(m_b m_s)$.

The  expressions of the isospin amplitudes can be rewritten  as follows:
\beqa
 a^u_{1/2}e^{i\delta_{1/2}}&=&\fa A_{\pi^+K^-}^u + \fb
A_{\pi^0\bar{K^0}}^u \nomb\\
             &=&r \lb \fa (a_1+a_4+a_{10}+(a_6+a_8)R_4)  
               + \sqrt{\frac{1}{6}} (a_4+a_6 R_5
-\frac{1}{2}(a_{10}+a_8R_5)) \right.\nomb\\
            && \left. - \sqrt{\frac{1}{6}}(a_2+\frac{3}{2}(a_9-a_7)X)  \rb \\
 a^c_{1/2}e^{i\delta_{1/2}}&=&\fa A_{\pi^+K^-}^c + \fb
A_{\pi^0\bar{K^0}}^c \nomb\\
&=& r\lb \fa(a_4+a_{10}+(a_6+a_8)R_4)
     + \sqrt{\frac{1}{6}} (a_4+a_6 R_5 -\frac{1}{2}(a_{10}+a_8R_5))
\right.\nomb\\ 
&& \left. - \sqrt{\frac{1}{2}\sqrt{\frac{3}{2}}}(a_9-a_7)X  \rb\\
 a^u_{3/2}e^{i\delta_{3/2}}&=&\fb A_{\pi^+K^-}^u - \fa
A_{\pi^0\bar{K^0}}^u \nomb\\
&=&r \fb \lb a_1+a_2X+\frac{3}{2}(a_9-a_7)X+\frac{3}{2}(a_{10}+a_8)R_4\rb \\
 a^c_{3/2}e^{i\delta_{3/2}}&=&\fb A_{\pi^+K^-}^c - \fa
A_{\pi^0\bar{K^0}}^c \nomb\\
&=&r \frac{\sqrt{3}}{2}(a_{10}+a_8R_4+(a_9-a_7)X)
\eeqa
and 
\beqa
\sqrt{\frac{2}{3}} b_{1/2}^c-\frac{1}{\sqrt{3}}a_{3/2}^c
&=&A_{\pi^-\bar{K^0}}^c \nomb\\
&=&r \lb a_4+a_6R_5-\frac{1}{2}(a_{10}+a_8R_5)\rb
\eeqa
where $r=\rk$ and  
$X=(f_\pi/f_K)(F^{BK}_0/F^{B\pi}_0)(m^2_B-m^2_K)/(m^2_B-m^2_\pi)$.
 In our numerical estimates, we will take $f_\pi=133$ MeV, $f_K=158$
MeV, $F_0^{B\pi}=0.36$ and
 $F_0^{BK}=0.41.$ There remains a large uncertainty in strange quark mass
$m_s$. For
 $m_s = (100 \sim 200)$ MeV, we find that the numerical values of those
amplitudes
are given by
\beqa
a_{1/2}^u &\simeq&  818\sim 846, \ a_{3/2}^u \simeq  709 \nomb \ , \\
a_{1/2}^c &\simeq& -( 103\sim 131), \ a_{3/2}^c \simeq  -7 \nomb  \ , \\
b_{1/2}^{c}&\simeq& -(72 \sim 100)\ .
\eeqa

\begin{figure}
\centerline{ \psfig{figure=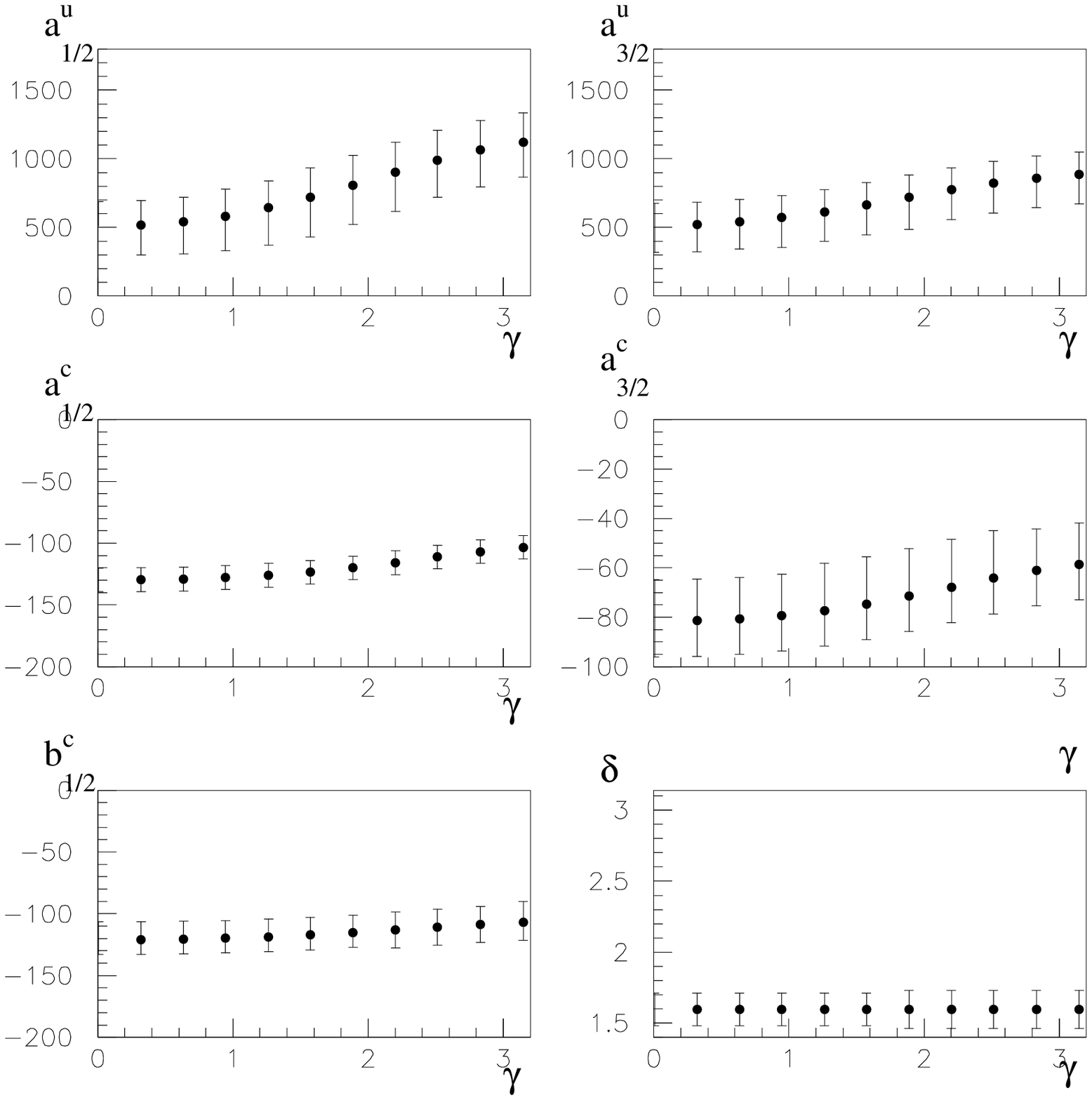, width=14cm} }
 \caption{The isospin amplitude fitted as a function of weak 
phase $\gamma$ from a $\chi^2-$ analysis of the recent CLEO data.
The vertical bars indicate the errors at 1$\sigma$ level. the strong phase
$\delta'$ is set to be zero. }
\label{iso00.eps}	
 \end{figure}

\begin{figure}
\centerline{ \psfig{figure=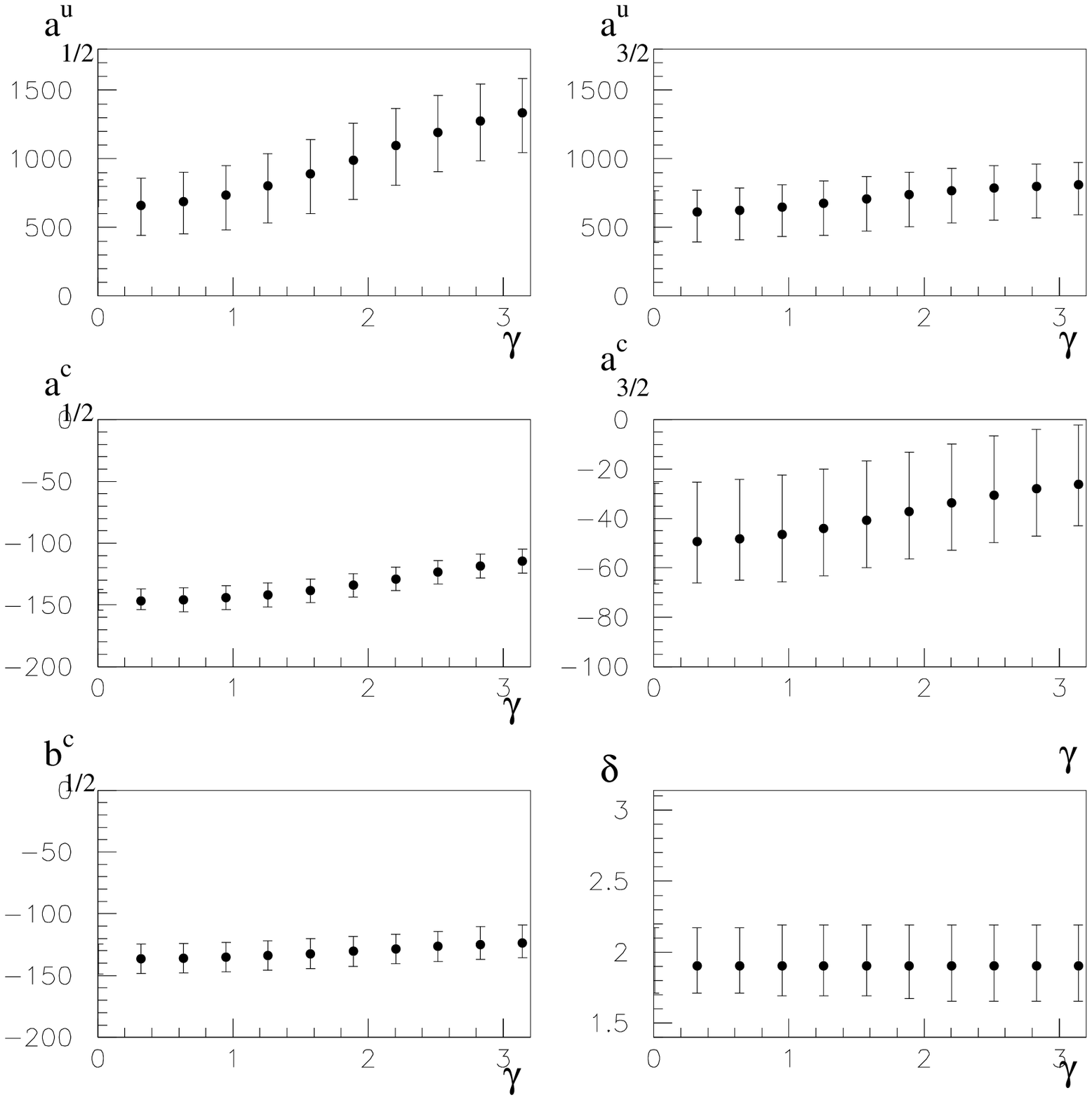, width=14cm} }
 \caption{The same as Fig.\ref{iso00.eps} but with $\delta'=\pi/6$ }
\label{iso30.eps}	
 \end{figure}

\begin{figure}
\centerline{ \psfig{figure=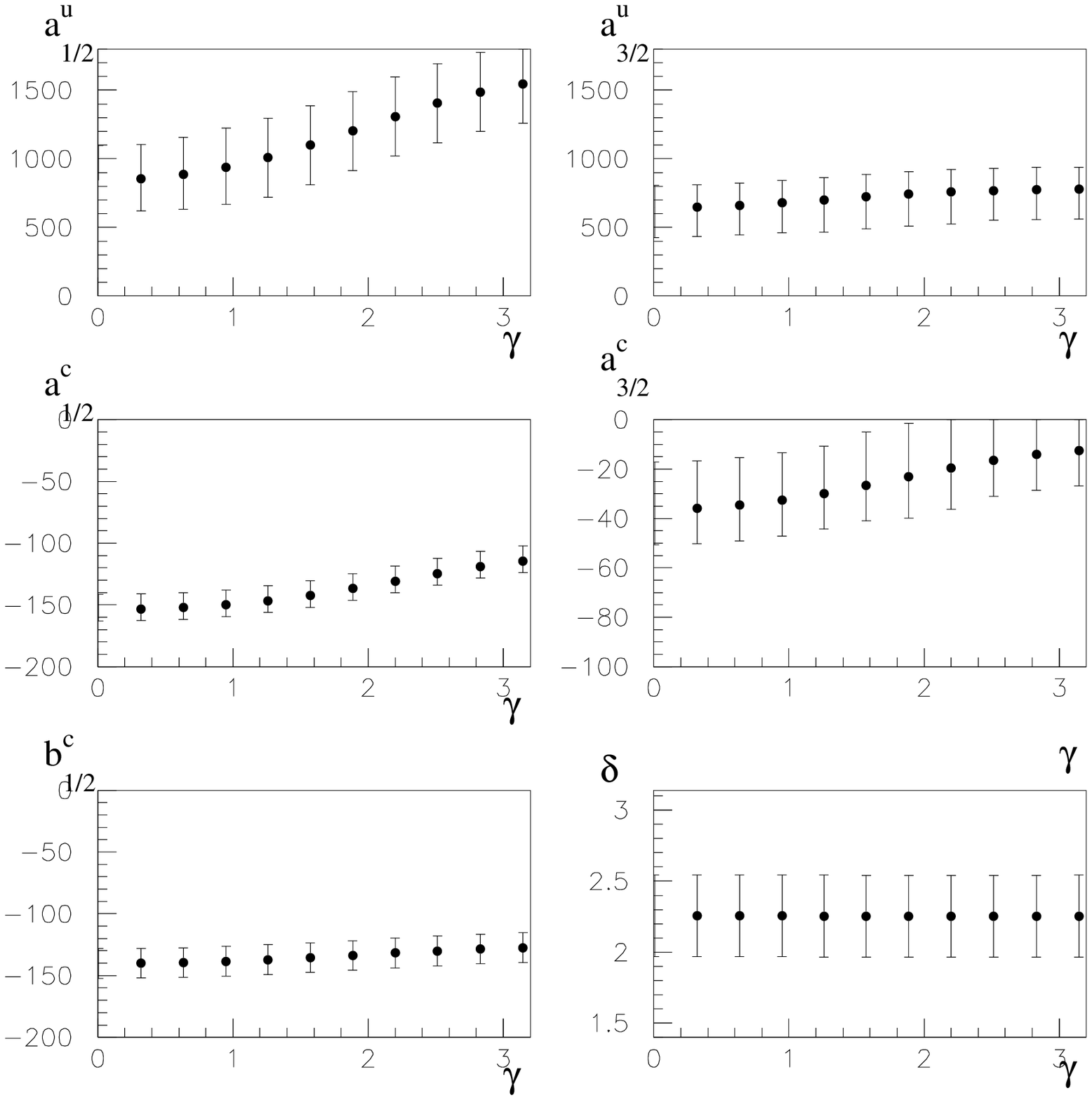, width=14cm} }
 \caption{ The same as Fig.\ref{iso00.eps} but with $\delta'=\pi/3$}
\label{iso60.eps}	
 \end{figure}

	The results from the $\chi^2$-fitting are shown in Fig.\ref{iso00.eps}-\ref{iso60.eps}, 
where the six amplitudes as well as their errors at 1$\sigma$ level are obtained 
as functions of the weak phase $\gamma$ with $\delta'$ fixed at $0,\pi/6,\pi/3$ respectively.
 The relative magnitudes of the amplitudes are consistent with the previous discussions.  
In our fit, the minimum value of $\chi^2$ is found to be extremly low
( typiclly  $\chi^2_{min}\sim 0.5\times 10^{-12}$). This means that the $\chi^2$-fits
are highly consistent and the 6 amplitudes are actually extracted as the solutions of Eqs.2-8.
It can be seen from the figures that the $\gamma$ dependence of $a_{1/2}^u$ is 
quite strong and the one of $a^{c}_{1/2}$ and $a^{u,c}_{3/2}$ is relatively weak. On the
other hand, it may be used to extract the angle $\gamma$ once one of those amplitudes 
can be determined or calculated in other independent ways. It is of interest to see that 
the $\gamma$ dependence of the amplitude $b^c_{1/2}$ and the strong phase
$\delta$ is weak, which shows that these two quantities are approximately fixed. 
The possibility of large $\delta$ was also suggested in Ref.\cite{hou} to explain 
the large branching ratio of the $B\to \pi^0 \bar{K^0}$ decay. Recently, perturbative QCD 
calculations have also shown a large strong phase\cite{lihn}. 
 
   In Fig.\ref{iso00.eps} where the phase difference $\delta'$ is set to be zero as usual, 
the $\chi^2$-fitting shows that the values of $a_{1/2}^u$ and $a_{3/2}^u$ may be comparable
with the ones from the theoretical estimations. only when the weak phase $\gamma$ is large. 
Especially for $\gamma > 2\pi/3$, the fitting values could coincide with the ones from 
factorization except for $a_{3/2}^c$. For $\gamma < 2\pi/3$, the two amplitudes are smaller than
the ones from naive factorization calculations. It appears that the factorization approach may 
become suitable for large weak phase $\gamma$. This phenomenon was observed
by most of the analyses in the literature which neglects the isospin amplitude $a_{3/2}^c$. 
 As a consequence, the resulting large value of $\gamma$ seems to be 
conflict with the one obtained from other constraints in the standard model. 
Before drawing the final conclusion, one also notice that in the factorization approach, 
one yields a zero strong phase $\delta=0^{\circ}$ which actually contradicts with the general 
fitting value $\delta \simeq \pm 95^{\circ}$. Therefore, the results of estimates 
 based on the naive factorization approach should be unreliable, and the
isospin amplitudes must receive additional large contributions. A large 
value for the relative strong phase $\delta  \pm 95^{\circ}$  implies
 that the final state interactions or inelastic rescattering effects must
be significant. 
		
\begin{figure}
\centerline{ \psfig{figure=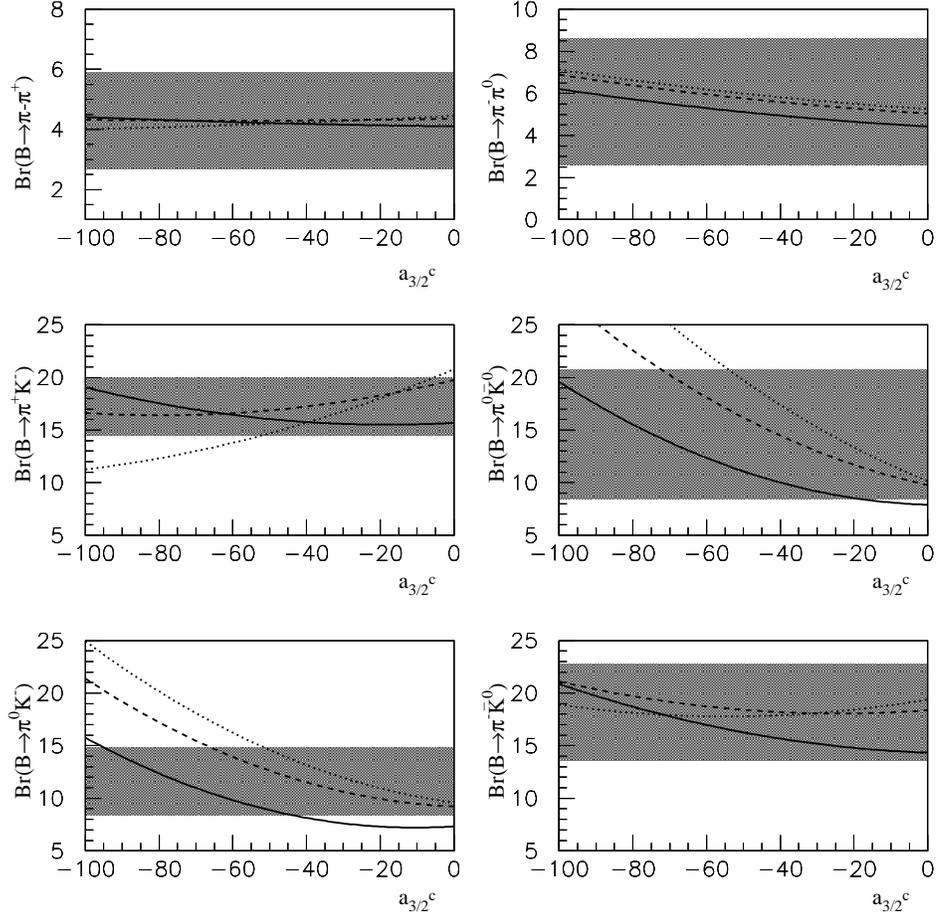, width=14cm} }
 \caption{The $a^c_{3/2}$ dependence of the 6 branching ratios( in units
of $10^{-6})$, for $\gamma=70^\circ$. The other parameters are at their 
central values.    The three curves in each plot
correspond to $\delta'=$ 0(solid), $\pi/6$ (dashed), $\pi/3$ (dot-dashed)
  The hashed bands indicate the errors(at 1$\sigma$) 
of the data. }
\label{a2c.eps}	
 \end{figure}

   We would like to stress that the most outstanding feature of the
$\chi^2$ analysis with $\delta'=0$ is that 
 the isospin amplitude $a^c_{3/2}$ is likely to be relatively larger  than
the one estimated from the naive factorization calculations. The fitting central value
of $a^c_{3/2}$ may be larger by a factor of $7\sim 9$. To explicitly see how 
the decay rates depend on  the isospin amplitude $a^c_{3/2}$, we plot
in Fig.\ref{a2c.eps} the 6 branching ratios of $B\to \pi\pi,\pi K$ decays
as functions of $a^c_{3/2}$. 
     It can be clearly seen that if $\delta'=0$,  a small value of
$a^c_{3/2} \sim -7$ is not able to reproduce all the CLEO data within the 
$1\sigma$ level, especially the data for the channels
of $\pi^0\bar{K^0}$ and $\pi^0 K^-$. To consistently describe the whole data, 
we need a relative large value $a^c_{3/2}\sim -75$ for fitting the central value of the data
which is about $10\%$ of the largest one $a_0^u$. 

      Within the standard model, it seems difficult to enhance the isospin 
amplitude $a^c_{3/2}$ by an order of magnitude even the inelastic FSI is
involved, this is because the main inelastic channels such as $B\to DD(DD_s,\eta_c K)\to \pi\pi(K)$,
only contribute to isospin-$\frac{1}{2}$ part of the decay amplitude.
In the SM, it is known that 
the ratio $a^c_{3/2}/a^u_{3/2}$ can be 
determined without the hadronic uncertainties in the flavor $SU(3)$ limit, this is because
the ratio only depends on the short distance Wilson coefficients\cite{neuberta3/2}. Thus
a large value of $a^c_{3/2}/a^u_{3/2}$ may indicate the existence of new 
physics. While all models beyond the standard model must effectively provide large contributions to the 
electroweak penguins in order to enhance the isospin amplitude $a^c_{3/2}$, 
such models are: SUSY with R parity violation, $Z'$-model and 
$Z$-mediated FCNC models, et. al.

 Let us now consider the case that $\delta'$ is non-zero, the situation then becomes
quite different. In Fig.\ref{iso30.eps} and Fig.\ref{iso60.eps}, it is seen
that $a^c_{3/2}$ decreases as the value of $\delta'$ increases. When
$\delta'$ reaches $\pi/3$ , $a^c_{3/2}$ will be consistent with the value yielded from 
factorization approach. It is also noticed from the figures that a large $\delta'$ leads to
large values of $a^u_{1/2}$ and $\delta$. The enhancement of $a^u_{1/2}$ may
be easily understood than the enhancement of $a^c_{3/2}$ due to final state interactions.
Nevertheless, as will be discussed below, the values of $\delta'$ cannot be too large 
due to the constraint of the upper bound of the branching ratio of $B \to \pi^0 \pi^0$.
     
\begin{figure}
\centerline{ \psfig{figure=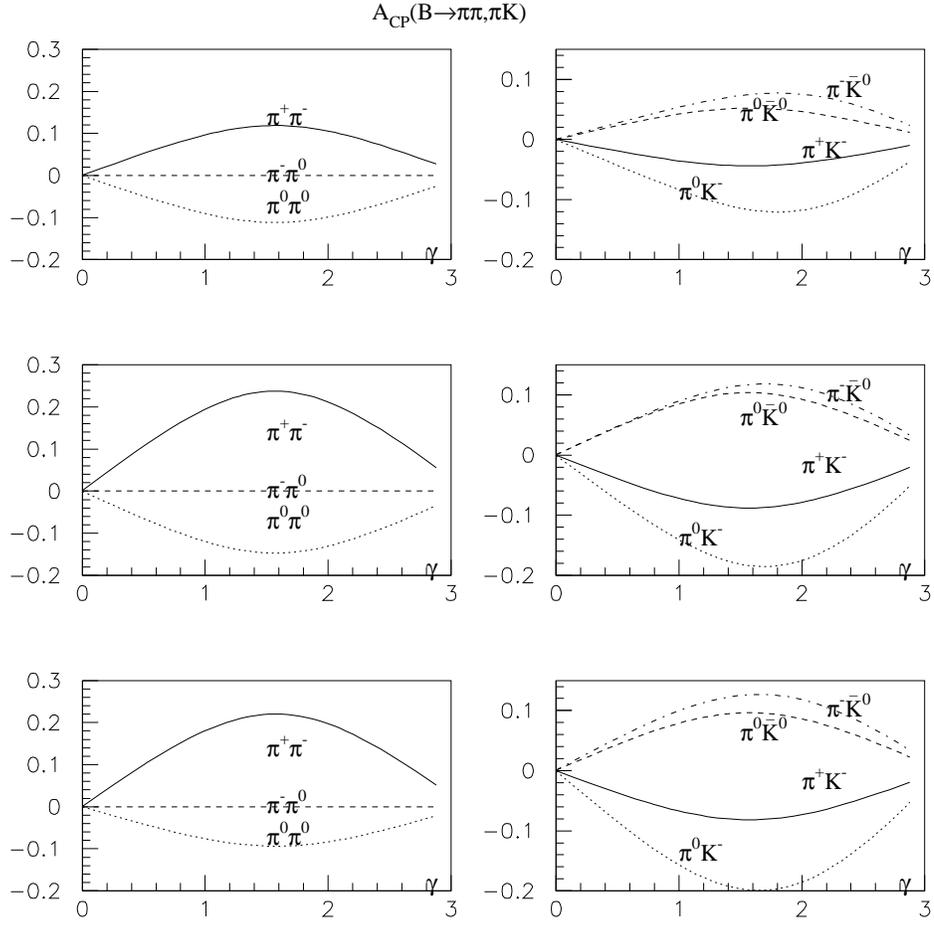, width=14cm} }
 \caption{The values of CP asymmetries $A_{CP}$ vs 
weak phase $\gamma$ with different $\delta'=0,\pi/6,\pi/3$ ( from up to down)
 The cures corresponding $|A_{CP}|$s for 
$\pi^0K^-$,$\pi^+\pi^-$,$\pi^0\bar{K^0}$,$\pi^+K^-$, $\pi^-\bar{K^0}$
and $\pi^-\pi^0$ are indicated in the plot.
}
\label{acp.eps}	
 \end{figure}

  When all the isospin amplitudes and strong phases are determined, one is
able to predict the direct CP asymmetries for all the relevant decay
channels. The direct CP asymmetry in $B\to \pi\pi,\pi K$ decays is defined in the
standard way
\beq
A_{CP} =\frac{\Gamma(\bar{B}\to \bar{f})-\Gamma(B\to f)}
             {\Gamma(\bar{B}\to \bar{f})+\Gamma(B\to f)} \equiv
a_{\epsilon''}^{f}
\eeq  
 where $f$ denotes the final state mesons. In Fig.\ref{acp.eps} we plot
several $|A_{CP}|$'s as functions of the weak phase $\gamma$ with different value 
of $\delta'$. When $\gamma$ is near $90^\circ$ and $\delta'=0$ one has 
$|A_{CP}(\pi^+K^-)|\sim 0.04$ which is in a good agreement with the most recent CLEO data 
$A_{CP} = -0.04 \pm 0.16$\cite{cleo-cp}. At this point, we have a reliable prediction 
for the direct CP violations in the following decay modes. The $A_{CP}$'s with
$45^\circ<\gamma<95^\circ$  read
\begin{eqnarray} 
 & & |A_{CP}(\pi^+K^-)|\simeq (2.5\sim 4)\% \ , \qquad
|A_{CP}(\pi^0\bar{K^0})|\simeq (2.5\sim 5)\% \ , 
\nonumber \\ 
 & & |A_{CP}(\pi^0K^-)|\simeq (7.5\sim 10) \%\ , \qquad
|A_{CP}(\pi^-\bar{K}^0)|\simeq (5 \sim 6.5)\% \ ,   
\nonumber \\
& & |A_{CP}(\pi^+\pi^-)|\simeq (7.5\sim 12.5)\% \ , 
\qquad |A_{CP}(\pi^0\pi^0)|\simeq (7.5\sim 12)\% \ .
\end{eqnarray}
for $\delta'= 0$, and 
\begin{eqnarray} 
 & & |A_{CP}(\pi^+K^-)|\simeq (5\sim 8)\% \ , \qquad
|A_{CP}(\pi^0\bar{K^0})|\simeq (8\sim 10)\% \ , 
\nonumber \\ 
 & & |A_{CP}(\pi^0K^-)|\simeq (12\sim 18) \%\ , \qquad
|A_{CP}(\pi^-\bar{K}^0)|\simeq ( 8\sim 12)\% \ ,   
\nonumber \\
& & |A_{CP}(\pi^+\pi^-)|\simeq (16\sim 24)\% \ , 
\qquad |A_{CP}(\pi^0\pi^0)|\simeq (10\sim 14)\% \ .
\end{eqnarray}
for $\delta'=30^\circ$. 
In spite of the large $\delta$, the smallness of CP asymmetries in 
$B\to \pi^-\pi^0$ decay is due to the absence 
of the interference between tree and penguin diagrams.
 
   There remains an unobserved decay mode in $B\to \pi\pi,\pi K$ decays,
which is the decay mode $B\to \pi^0\pi^0$ ( the CLEO collaboration has already reported the
indication of $Br(B\to \pi^0\pi^-) \simeq 5.6 $ ). As all the relevant isospin amplitudes have
been determined as functions of $\gamma$, it allows us to predict $B\to \pi^0\pi^0$ as a
function of $\gamma$. It is interesting to note that our $\chi^2$-analysis shows that the
resulting branching ratio $Br(B\to \pi^0\pi^0)$ is almost independent of the weak phase $\gamma$.
Its value at $\delta'=0$ is close to the one of $B\to \pi^+\pi^-$ decay
\begin{equation} 
Br(B\to \pi^0\pi^0)\sim 4.6 \times 10^{-6}\,.
\end{equation}
with $\delta'$ increasing, the branching ratio becomes larger and can reach $\sim 7( 10)\times 10^{-6}$
when $\delta'=30^\circ( 60^\circ)$.  Such a large branching ratio is about an order of magnitude larger 
than the prediction based  on factorization calculations. Most recently, the CLEO collaboration
reported an upper bound of $Br(B\to \pi^0\pi^0) < 5.7 \times 10^{(-6)}$\cite{pi0data}, 
this will impose a strong constraint on the value of $\delta'$ (see Fig. 6). It is seen that
to be consistent with the data at the $1\sigma$ level, the upper bound of 
the branching ratio $Br(B\to \pi^0\pi^0)$ limits $\delta'$ to be less than $\sim 50^\circ$.  

\section{conclusions}
   In summary, we have made a general less model-dependent
investigation on the charmless B-meson decays by using the $\chi^2$ analysis 
based on the most recent CLEO data. We have used the most general isospin decomposition
with two independent strong phases.   All the isospin amplitudes in rare hadronic 
$B$ decays $B\to \pi\pi,\pi K$ can be determined as functions of the weak phase 
$\gamma$ and one strong phase $\delta'$. 
The effects of two equal and unequal strong phases are studied in detail.
It is found that
the isospin amplitude $b^c_{1/2}$ and the strong phase $\delta$ only slightly depend on 
the phase $\gamma$.
An important observation under the equal strong phase assumption
is the relative large isospin amplitude $a^c_{3/2}$, 
$(a_{3/2}^c \simeq -70)$ where the central value is about 5 times greater than the 
one obtained from the factorization calculations. When the two strong phases are
not equal, the allowed values of $a^c_{3/2}$ decrease as their difference, i.e., $\delta'$,
increases. For the most general case with two rather than one large FSI strong phases, the magnitude of 
all the isospin amplitudes may be around the one estimated from the factorization approach. 
Nevertheless, one needs to find out the mechanism of producing large strong phases. This could directly 
be tested by measuring the branching ratio $Br(B\to \pi^0\pi^0)$. 

  The direct CP asymmetries $A_{CP}^f$ for all the relevant decay channels
have also been given as functions of $\gamma$. The resulting numerical value for
$A_{CP}^{\pi^+K^-}$ is consistent with the most recent data. When taking $\gamma$ to be 
in a reasonable range $\gamma = 45^{\circ}\sim 95^{\circ}$, we are led to the results
given in eqs.(35) and (36), which can be directly tested by experiments in the near future. 
 A resulting large branching ratio $Br(B\to \pi^0\pi^0)$ which is comparable with
the one $Br(B\to \pi^+\pi^-)$ will also provide an important and consistent test.

   From the most general analysis presented in this paper, the data appears to
strongly suggest that final state interactions and inelastic rescattering 
effects must be significant and play an important rule in the charmless 
$B\to \pi\pi,\pi K$ decays. Otherwise, our general analyses may be interpreted as hinting
at the existence of 
new physics. For a more definite conclusion, one needs more precise data. 
The two B-factories BaBaR and BELLE are expected to provide us with more information from the charmless
decays.

{\bf Acknowledgments:} 

 The author (C.Q.G.) was supported by the NSC under contract 
number NSC 89-2112-M-007-013. J.N.N. is partially supported by the Natural Science 
and Engineering Council of Canada. 
Two of us (Y.L.W. and Y.F.Z.) were supported in part by NSF of China under grant 
No. $\#$ 19625514. Y.L.W acknowledges the CTS for the support during his visit and  
he would also like to thank H.Y. Cheng, X.G. He, W.S. Hou and H.N. Li for useful discussions.

%\begin{figure}
%\centerline{ \psfig{figure=del2.eps, width=14cm} }
% \caption{}
%\label{del2.eps}	
% \end{figure}

\begin{figure}
\centerline{ \psfig{figure=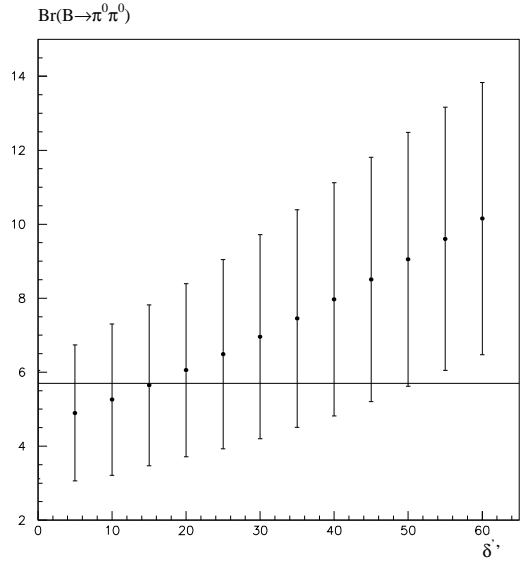, width=8cm} }
 \caption{the branching ratio of $B\to \pi^0\pi^0$ (in units of $10^{-6}$ predicted as
 function of $\delta'$(in degree). the solid line indicate the
 upper bound observed reported by CLEO collaboration.}
\label{pi0err.eps} 
\end{figure}

\end{document}